# Identification of Effective Connectivity Subregions

Ruben Sanchez-Romero, Joseph D. Ramsey, Kun Zhang, Clark Glymour

Department of Philosophy, Carnegie Mellon University, Pittsburgh, Pa., USA

July 23, 2019

## Abstract

Standard fMRI connectivity analyses depend on aggregating the time series of individual voxels within regions of interest (ROIs). In certain cases, this spatial aggregation implies a loss of valuable functional and anatomical information about smaller subsets of voxels that drive the ROI level connectivity. We use two recently published graphical search methods to identify subsets of voxels that are highly responsible for the connectivity between larger ROIs. To illustrate the procedure, we apply both methods to longitudinal high-resolution resting state fMRI data from regions in the medial temporal lobe from a single individual. Both methods recovered similar subsets of voxels within larger ROIs of entorhinal cortex and hippocampus subfields that also show spatial consistency across different scanning sessions and across hemispheres. In contrast to standard functional connectivity methods, both algorithms applied here are robust against false positive connections produced by common causes and indirect paths (in contrast to Pearson's correlation) and common effect conditioning (in contrast to partial correlation based approaches). These algorithms allow for identification of subregions of voxels driving the connectivity between regions of interest, recovering valuable anatomical and functional information that is lost when ROIs are aggregated. Both methods are specially suited for voxelwise connectivity research, given their running times and scalability to big data problems.

## Introduction

The high dimensionality of functional magnetic resonance BOLD signals measured at the voxel level is typically reduced by aggregating some of the voxels into regions of interest (ROIs) consisting of hundreds or sometimes thousands voxels and combining the signals from voxels within a cluster into a single variable. Whatever the clustering method, whether anatomical or statistical, it seems unlikely that voxels within a large cluster are homogeneous in their functions[1], specifically in their signaling patterns. Subsets of voxels internal to a ROI that are specialized for communicating with other ROIs provide an opportunity to search for intra-ROI connectivity differentiation.

This problem, and the larger one of identifying common patterns of signaling connections throughout the cortex, has been addressed by "connectivity-based parcellation" (Eickhoff et al., 2015). The idea is to compute the time series correlation of every voxel with the rest of the voxels in the cortex (Anteraper et al., 2018) or the time series correlations of voxels in some smaller region such as the insula with the rest of the voxels in the cortex (Kelly et al., 2012), and then cluster the voxels with similar patterns of connectivity. In contrast, the method proposed here assumes that a set of communicating regions of interest have been identified and seeks for each pair of ROIs in the set, those voxels within them that drive their signaling connections. *Figure 1* illustrates this idea with an arbitrary model at the region of interest level and its corresponding model at the subregion level. Using the non-Gaussian distribution of the BOLD signal under the linearity assumption, the aim is to detect functionally distinct subregions inside larger regions while avoiding the risks of false positive connections expected from correlation and partial correlation approaches.

We apply two recently published procedures, the FASK and Two-Step algorithms (Sanchez-Romero et al., 2019), to infer voxel level connectivity using resting state fMRI data from the medial temporal lobe, for which there is anatomical, histological, and experimental work distinguishing neural regions and axonal connectivities (Lavenex et al., 2000; Insausti et al., 2004) The results of the two search procedure are used to identify "connectivity subregions" of ROIs in the medial temporal lobe

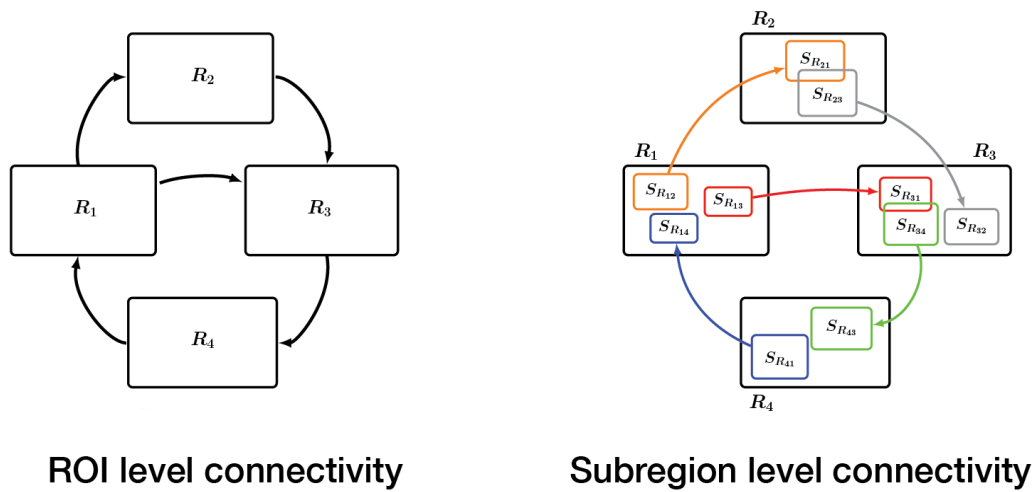

ROI level connectivity      Subregion level connectivity

*Figure 1.* Illustrative arbitrary connectivity model at the region of interest level (left) and a parcellation into connectivity subregions driving the larger level connectivity (right).

---

[1] By "function" we refer to the mechanisms that produce the associations of voxel time series. We do not use "function" or "functional" to mean "correlation" or "correlational."

## Methods

We use the directed graph framework (Spirtes, et al., 2000; Pearl, Glymour and Jewel, 2017) to represent causal relations among variables. In our application, the variables are the time series of voxel BOLD signals. The inference task is to correctly identify from the BOLD time series a directed graph that represents communication relationships among these variables. Each directed edge in such a graph represents an influence from one variable to another that does not pass through any other measured variable. A sequence of edges in the same direction indicates a communication pathway. A pair of paths from one variable, *X*, to each of two other variables *Y* and *Z*, represents a common cause in *X*, that produces a statistical association of the *Y* and *Z* signals, whether or not *Y* and *Z* directly signal one another. Since communication in the brain is often reciprocal, the graphical representation allows cyclic graphs.

The directed graph representation illuminates systematic flaws in search methods that interpret correlations or partial correlations of BOLD time series as direct communication relations. In the absence of any direct connection between them, two voxels can be correlated by an indirect path between them or by a third voxel that signals both. Partial correlation will associate any pair of voxels that have no direct or indirect signaling connection, provided that there is at least one other voxel signaled by both. Both correlation and partial correlation methods are therefore liable to yield false connections even given the true probability distributions or in the limit as the sample size increases without bound. In theory, the FASK and Two-Step algorithms, which assume linear causal relations and exploit non-Gaussian features of BOLD time series, avoid these problems.

*The FASK algorithm*

The FASK algorithm first estimates which time series voxels are adjacent in a graphical representation, i.e, the direction or directions of influence are unspecified. The procedure uses the first stage of the PC-stable algorithm (referred here as FAS-stable) (Colombo and Maathuis, 2014), which exploits a series of iterative conditional independence assessments. The procedure avoids positing adjacencies between voxels connected only through intermediate voxels, and avoids adjacencies between voxels related only by a common cause or voxels related only by influencing shared voxels. Conditional independence is assessed by the Bayes Information Criterion (BIC) score (Schwarz, 1978) with an extra penalty discount *c* specified by the user. Each of the $X_i$ — $X_j$ inferred adjacencies will be oriented as a 2-cycle $X_i \leftrightarrow X_j$, or as $X_i \rightarrow X_j$, or $X_i \leftarrow X_j$.

The correctness of the output of FASK assumes that the noise terms in the causal process are skewed. The 2-cycle detection rule is applied by testing $corr(X_i, X_j) \neq corr(X_i, X_j | X_i > 0)$, and $corr(X_i, X_j) \neq corr(X_i, X_j | X_j > 0)$. If both inequalities hold, the edges $X_i \rightarrow X_j$ and $X_i \leftarrow X_j$ are added to the output graph. If no 2-cycle is inferred, the Left-Right orientation rule is applied: Orient $X_i \rightarrow X_j$ if: $\{ E(X_i X_j | X_i > 0) / sqrt[E(X_i^2 | X_i > 0) E(X_j^2 | X_i > 0)] \} > \{ E(X_i X_j | X_j > 0) / sqrt[E(X_i^2 | X_j > 0) E(X_j^2 | X_j > 0)] \}$; otherwise orient $X_i \leftarrow X_j$. FASK orients all inferred adjacencies returning a fully oriented graph as output.

For some models, where the true coefficients of a 2-cycle between $X_i$ and $X_j$ are more or less equal in magnitude but opposite in sign, FAS-stable may fail to detect an adjacency between $X_i$ and $X_j$ when in fact a 2-cycle exists. In these cases, FASK checks explicitly whether $abs(corr(X_i, X_j | X_i > 0) - corr(X_i, X_j | X_j > 0)) > \Delta$, where $\Delta$ is a pre-specified positive value. If so, the adjacency is added and oriented using the aforementioned rules. The idea for the heuristic extra adjacency rule is that if $X_i$ and $X_j$ are independent, then $abs(corr(X_i, X_j | X_i > 0) - corr(X_i, X_j | X_j > 0)) = 0$, so if this difference is not zero, there must be an indirect path or a common cause between $X_i$ and $X_j$. If the difference is very large, a very short indirect path is inferred and so an adjacency is added. The $\Delta$ threshold is a parameter than can be adjusted. A value of $\Delta = 0.3$ gave good results in the cyclic simulations of Sanchez-Romero et al., (2019), and is used here.

Proof of correctness, performance under simulations and pseudo-code for FASK are reported in Sanchez-Romero et al. (2019). The FASK algorithm is implemented in the TETRAD open software, available as a graphical user interface at www.ccd.pitt.edu/tools/ or as source code at github.com/cmu-phil/tetrad.

*The Two-Step Algorithm*

The Two-Step algorithm (Sanchez-Romero et al., 2019), represents a linear causal structure, with possible unmeasured confounding variables and cyclic relationships in matrix form as: $X = BX + MC + E$. Where X are the observed variables; B is a directed connectivity matrix of linear coefficients defining the causal structure between observed variables; M is a matrix indicating which observed variables (rows) are affected by unmeasured confounders (columns); C are the unmeasured confounders, if they exist; and E are the mutually independent noise or disturbance terms. Two-Step applies the principles of independent component analysis (ICA) (Hyvärinen and Oja, 2000) by expressing the observed variables X as a mixture of a set of unobserved components: $X = (I - B)^{-1}(MC + E)$; where I is the identity matrix; $(I - B)^{-1}$ defines the mixing matrix of the unobserved components; and $(MC + E)$ defines the unobserved components. In our application, we ignore the possible unobserved confounders C. This results in the model $X = (I - B)^{-1}E$, which can be expressed as $E = (I - B)X$.

The Two-Step algorithm uses standard Independent Components Analysis (ICA) methods to estimate the directed connectivity matrix of coefficients B in a divide-and-conquer manner. In the first step the algorithm infers the undirected adjacencies between the observed variables in X, and makes use of these adjacencies to reduce the number of free-parameters in the estimation of (I - B). In a second step, the algorithm optimizes the non-zero entries of (I - B) to make the components in E as independent or non-Gaussian as possible, with a sparsity constraint on B.

Step one of the Two-Step algorithm can use the FAS-stable procedure or adaptive lasso (Zou, 2006) to find the undirected adjacencies over the variables. Adaptive lasso is a regression method with adapted penalization for each coefficient according to an initial ordinary least squares multiple regression. We use adaptive lasso in this application, since Sanchez-Romero et al., (2019) reported better performance of Two-Step under simulations when using this algorithm in step one.

Step two of the Two-Step algorithm estimates the free parameters in the matrix (I - B) that maximize the independence or non-Gaussianity of the estimated components E, with a sparsity constraint on the free parameters of (I - B). The current implementation of the algorithm uses ICA with sparse connections achieved by adaptive lasso (Zhang et al., 2009), and thus the estimated (I - B) matrix entries are expected to be as small as possible. As a consequence of this sparsity constraint and the initialization of the free parameters of (I - B) with small values, in the presence of a cyclic structure, the Two-Step algorithm tends to find the most stable solution, characterized by small connectivity coefficients in the cycles and all eigenvalues of the estimated B matrix of coefficients less than one in absolute value.

The Two-Step algorithm returns an estimation of the components E, and a matrix of estimated coefficients B, which encodes the connectivity between observed variables X. In practice, the estimated matrix B can be further thresholded to censor values that are very close to zero, and a directed, possibly cyclic graph can be reconstructed from it.

Two-Step uses a penalty of $(1/\lambda)log(n)$ for the (I - B) matrix estimation in step two, where *n* is the number of datapoints and $\lambda > 0$ is adjusted by the user; and a zeroing threshold *t* for the absolute values of the estimated matrix of coefficients B, such that if $abs(B_{ij}) < t$, then $B_{ij} = 0$. In step one of Two-Step, adaptive lasso uses a penalty term $\sigma > 0$, also adjusted by the user. Performance of Two-Step under cyclic simulations is reported in Sanchez-Romero et al., (2019), and a Matlab implementation is available at http://github.com/cabal-cmu/Two-Step/.

**Data**

We used longitudinal single subject resting state data presented in Poldrack et al., (2015) and Laumann et al., (2015) and freely available as raw scans at

https://openneuro.org/datasets/ds000031/versions/00001. Russell Poldrack kindly provided functional preprocessed data at the voxel level for regions of interest defined bilaterally in the medial temporal lobe. The data came from one single individual resting state fMRI longitudinal dataset, acquired with a 3T MRI scanner, voxel size = 2.4 mm isotropic, TR = 1.16 sec and 10 minutes scan length producing 518 volumes (datapoints). Data were collected during 18 months, resulting in 80 datasets. Functional data were preprocessed with intensity normalization, motion correction, atlas transformation, distortion correction using a mean field map, and resampling to 2 mm atlas space. 57 volumes were removed due to drifting, resulting in 461 data points. T2-weighted anatomical data were collected using a T2-SPACE sequence (sagittal, 256 slices, 0.8 mm isotropic resolution, TE = 565 ms, TR = 3200 ms, PAT = 2, 8:24 min scan time). Full information about acquisition is reported in Poldrack et al., (2015) and Laumann et al., (2015).

Regions of interest in the medial temporal lobe were defined manually by Jackson C. Liang according to procedures established by the Preston Laboratory at The University of Texas at Austin (Liang et al., 2012). The high-resolution T2 anatomical images allowed a more reliable delineation of hippocampal subfields in the body, head and tail of the hippocampus; Insausti et al., (2004) and Duvernoy et al., (2013) were used as anatomical guidelines. The dentate gyrus (DG) is hardly separable in functional MRI from cornu ammonis 3 (CA3) and cornu ammonis 2 (CA2), resulting in the comprehensive ROI labeled CA32DG (Zeineh et al., 2017; Ekstrom et al., 2009; Preston et al., 2010). The entorhinal cortex is also a complex structure for which several subregions have been distinguished in the literature but are not distinguished in our data (Kerr et al., 2007; Canto et al., 2012). The final anatomical parcellation comprised four regions of interest defined bilaterally for subiculum (SUB), cornu ammonis 1 (CA1), CA32DG, and entorhinal cortex (ENT).

The availability of multiple resting state scans for the same individual permits multiplying the sample size by appending scans, while reducing concerns about co-alignment and false positive associations due to mixing of distributions from scans of different individuals. This strategy is used here to construct eight datasets for each hemisphere by concatenating, without replacement, ten temporally ordered individual sessions (4610 data points). These eight datasets are used to measure the consistency of the results across time for this individual.

**Defining Connectivity Subregions**

To obtain connectivity subregions for a set of pairs of regions, a voxel level graph comprising all the voxels of all the regions considered is first inferred, then subgraphs for the voxels of each pair of regions are extracted from this graph, finally subregions for each pair of regions are obtained from the subgraphs. For clarity, this idea is schematized below as *Connectivity*

*Subregions Algorithm*, and applied to the medial temporal lobe data in the following subsections.

---

Connectivity Subregions Algorithm

---

**Input**: A dataset $D$ defined over a set of voxels $V = \{V_1 \cup \ldots \cup V_R\}$ for $R$ regions of interest, where $V_i$ is the set of exclusive voxels of region $i$; and a connectivity search method $M$.
**Output**: Two connectivity subregions for each pair of regions of interest.

1:     Infer a voxel level directed graph $G = <V, E>$ applying $M$ to $D$
2:     Convert $G$ to an adjacency graph $G_A = <V, A>$ by disregarding edges $E$ orientations
3:     **for** each pair of regions $P$, $Q$ of interest **do**:
4:         from $G_A$ extract subgraph $G_{P,Q} = <V_P \cup V_Q, A_{PQ}>$, for which $A_{PQ}$ is the set of adjacencies in $G_A$ between voxels $v_P \in V_P$ and voxels $v_P \in V_P$
5:         return connectivity subregions for $P$ and $Q$ as the sets: $S_P = \{v_P \in V_P \mid$ the adjacency $v_P - v_Q$ exists in $A_{PQ}\}$ and $S_Q = \{v_Q \in V_Q \mid$ the adjacency $v_P - v_Q$ exists in $A_{PQ}\}$
6:     **end for**

---

*Voxel level graph results: consistency across datasets*

FASK and Two-Step were run separately on each of the eight concatenated datasets for the left hemisphere (results for the right hemisphere are similar and included in Supplementary Material). Each dataset consists of 570 voxels (variables) comprising a set of four different regions of interest: ENT (141 voxels), CA32DG (178 voxels), CA1 (153 voxels) and SUB (98 voxels), with 4610 data points. Inclusion of all the voxels in the dataset is important to reduce false connections region to region connections due to confounding voxels (common causes of activities of two or more voxels in distinct regions of interest), or intermediate voxels in a causal chain between voxels in two regions passing through a third region. FASK was run with penalty $c = 1$ for the BIC score of FAS-stable, *alpha* = $10^{-7}$ for the 2-cycle detection tests of difference of correlations and $\Delta = 0.3$ for the extra adjacency heuristic rule. To match the sparsity obtained with FASK, Two-Step was run with $\sigma = 75log(n)$ for the penalty of adaptive lasso in step one, $\lambda = 20$ for the penalty of the sparse ICA estimation, and zeroing threshold of $t = 0.05$ for *abs*(B).

To quantify the consistency of the results across the eight temporally ordered datasets, the graph obtained for each of the datasets is compared against each of the other seven graphs. The comparison is measured with the Jaccard similarity index (Jaccard et al., 1912), which calculates the number of edges in common (intersection) between two graphs as a proportion of the total number (union) of edges in the two graphs. The Jaccard index is between 0 and 1, where 1 means that both graphs are identical. It is important to consider that the Jaccard index is a measure of consistency, not informativeness. An overfitting method that produced a complete undirected graph for every dataset will have a Jaccard similarity index of 1 for every comparison.

Formally, consider graphs $G_1$ and $G_2$, which were inferred from two distinct datasets $D_1$ and $D_2$. The Jaccard index is obtained by the function, $J(E_{G1}, E_{G2}) = |E_{G1} \cap E_{G1}| / |E_{G1} \cup E_{G2}|$, where $E_{G1}$ and $E_{G2}$ are the set of edges of graphs $G_1$ and $G_2$, and $|\cdot|$ indicates the number of elements in the set. This index can be defined for directed or undirected graphs. *Table 1* and *Table 2* show Jaccard similarity indices for undirected graphs (the output graphs without taking into account the orientations of the edges) and directed graphs across the eight concatenated datasets, for FASK and Two-Step, respectively. For both algorithms, the Jaccard index across all the undirected graphs is similar and around 0.7. For the directed graphs, the index is also similar across graphs but smaller around 0.3, showing lower inter-session consistency in the orientations than in the adjacencies. It is possible that consistency would increase if stronger penalizations were used, so that the methods would return only the strongest connections. It is also worth noting that the connectivity might change to some extent across different datasets which were collected during different time periods.

Undirected Graphs

| Datasets | 1: (01, 10) | 2: (11, 20) | 3: (21, 30) | 4: (31, 40) | 5: (41, 50) | 6: (51, 60) | 7: (61, 70) | 8: (71, 80) |
|---|---|---|---|---|---|---|---|---|
| 1: (01, 10) | 1 | 0.706 | 0.691 | 0.692 | 0.687 | 0.682 | 0.681 | 0.699 |
| 2: (11, 20) |   | 1 | 0.714 | 0.708 | 0.723 | 0.707 | 0.715 | 0.733 |
| 3: (21, 30) |   |   | 1 | 0.7 | 0.689 | 0.707 | 0.69 | 0.729 |
| 4: (31, 40) |   |   |   | 1 | 0.701 | 0.687 | 0.693 | 0.713 |
| 5: (41, 50) |   |   |   |   | 1 | 0.691 | 0.697 | 0.703 |
| 6: (51, 60) |   |   |   |   |   | 1 | 0.67 | 0.709 |
| 7: (61, 70) |   |   |   |   |   |   | 1 | 0.691 |
| 8: (71, 80) |   |   |   |   |   |   |   | 1 |

Directed Graphs

| Datasets | 1: (01, 10) | 2: (11, 20) | 3: (21, 30) | 4: (31, 40) | 5: (41, 50) | 6: (51, 60) | 7: (61, 70) | 8: (71, 80) |
|---|---|---|---|---|---|---|---|---|
| 1: (01, 10) | 1 | 0.299 | 0.321 | 0.308 | 0.314 | 0.301 | 0.297 | 0.294 |
| 2: (11, 20) |   | 1 | 0.33 | 0.311 | 0.365 | 0.326 | 0.324 | 0.31 |
| 3: (21, 30) |   |   | 1 | 0.328 | 0.333 | 0.325 | 0.319 | 0.298 |
| 4: (31, 40) |   |   |   | 1 | 0.327 | 0.314 | 0.325 | 0.311 |
| 5: (41, 50) |   |   |   |   | 1 | 0.321 | 0.322 | 0.306 |
| 6: (51, 60) |   |   |   |   |   | 1 | 0.319 | 0.308 |
| 7: (61, 70) |   |   |   |   |   |   | 1 | 0.291 |
| 8: (71, 80) |   |   |   |   |   |   |   | 1 |

*Table 1*. **Jaccard similarity index for FASK** voxel level graphs from left medial temporal lobe data, one individual sampled across 80 sessions. Results indicate the Jaccard index for each pair of datasets, separately for undirected graphs and directed graphs. The Jaccard similarity index goes from 0 to 1, where 1 indicates complete similarity. Each dataset consists of 570 voxels corresponding to four regions of interest in the left hemisphere medial temporal lobe and ten temporally ordered sessions concatenated, resulting in 4610 datapoints. Each row/column indicates in parenthesis the range of the sessions that were concatenated for that particular dataset.

### Undirected Graphs

| Datasets | 1: (01, 10) | 2: (11, 20) | 3: (21, 30) | 4: (31, 40) | 5: (41, 50) | 6: (51, 60) | 7: (61, 70) | 8: (71, 80) |
|---|---|---|---|---|---|---|---|---|
| 1: (01, 10) | 1 | 0.708 | 0.684 | 0.683 | 0.678 | 0.685 | 0.631 | 0.7 |
| 2: (11, 20) |   | 1 | 0.696 | 0.714 | 0.726 | 0.689 | 0.702 | 0.711 |
| 3: (21, 30) |   |   | 1 | 0.712 | 0.675 | 0.687 | 0.659 | 0.682 |
| 4: (31, 40) |   |   |   | 1 | 0.679 | 0.69 | 0.663 | 0.688 |
| 5: (41, 50) |   |   |   |   | 1 | 0.669 | 0.678 | 0.687 |
| 6: (51, 60) |   |   |   |   |   | 1 | 0.639 | 0.699 |
| 7: (61, 70) |   |   |   |   |   |   | 1 | 0.648 |
| 8: (71, 80) |   |   |   |   |   |   |   | 1 |

### Directed Graphs

| Datasets | 1: (01, 10) | 2: (11, 20) | 3: (21, 30) | 4: (31, 40) | 5: (41, 50) | 6: (51, 60) | 7: (61, 70) | 8: (71, 80) |
|---|---|---|---|---|---|---|---|---|
| 1: (01, 10) | 1 | 0.393 | 0.388 | 0.394 | 0.39 | 0.393 | 0.351 | 0.375 |
| 2: (11, 20) |   | 1 | 0.382 | 0.411 | 0.414 | 0.392 | 0.386 | 0.368 |
| 3: (21, 30) |   |   | 1 | 0.411 | 0.368 | 0.391 | 0.372 | 0.372 |
| 4: (31, 40) |   |   |   | 1 | 0.386 | 0.39 | 0.363 | 0.363 |
| 5: (41, 50) |   |   |   |   | 1 | 0.383 | 0.389 | 0.368 |
| 6: (51, 60) |   |   |   |   |   | 1 | 0.357 | 0.382 |
| 7: (61, 70) |   |   |   |   |   |   | 1 | 0.357 |
| 8: (71, 80) |   |   |   |   |   |   |   | 1 |

*Table 2*. **Jaccard similarity index for Two-Step** voxel level graphs from left medial temporal lobe data, one individual sampled across 80 sessions. Results indicate the Jaccard index for each pair of datasets, separately for undirected graphs and directed graphs. The Jaccard similarity index goes from 0 to 1, where 1 indicates complete similarity. Each dataset consists of 570 voxels corresponding to four regions of interest in the left hemisphere medial temporal lobe and ten temporally ordered sessions concatenated, resulting in 4610 datapoints. Each row/column indicates in parenthesis the range of the sessions that were concatenated for that particular dataset.

*Subgraphs for pairs of regions: consistency across datasets*

From the full voxel level graph estimated above (570 voxels encompassing four regions of interest), it is possible to extract voxel level subgraphs for each pair of regions for which we aim to identify communication subregions. Properly, to obtain communication subregions we only need adjacency subgraphs, since subregions only indicate voxels that are driving the communication, irrespectively of the direction, between two larger regions of interest.

For a voxel level adjacency graph $G_A = <V, A>$, with set of adjacencies $A$ and set of voxels $V = \{V_1 \cup ... \cup V_R\}$ for $R$ regions of interest and $V_i$ the set of exclusive voxels of region $i$, define an adjacency subgraph for two regions $P$ and $Q$ as, $G_{P,Q} = <V_P \cup V_Q, A_{PQ}>$, for which $V_P$, $V_Q \subset V$, $A_{PQ} \subset A$. $V_P$ is the exclusive set of voxels of region $P$, $V_Q$ the exclusive set of voxels of region $Q$, and $A_{PQ}$ is the set of adjacencies between voxels of the form $v_P — v_Q$, for $v_P \in V_P$ and $v_Q \in V_Q$. In other words, the subgraph $G_{P,Q}$ is composed exclusively of adjacencies between voxels in regions $P$ and $Q$ with no intra-region adjacencies.

Following the standard connectivity model for the medial temporal lobe (Lavenex et al., 2000), the next five pairs of regions were selected to extract five subgraphs: (*ENT, CA32DG*), (*CA32DG, CA1*), (*CA1, SUB*), (*SUB, ENT*) and (*ENT, CA1*). As in the previous section, the Jaccard index is used to measure the consistency of the resulting subgraphs across the eight datasets. *Table 3* shows mean Jaccard indices for each subgraph across the eight datasets, for FASK and Two-Step. These results show lower consistency for the subgraphs $G_{ENT, CA32DG}$ and $G_{ENT, CA1}$. These are sparser subgraphs possibly reflecting a small number of active functional connections during rest or connectivities that are difficult to capture with these data, or methods, or greater actual variations in connectivities at different scanning times, or all of the above.

| FASK | | Two-Step | |
|---|---|---|---|
| $G_{ENT, CA32DG}$ | 0.06 (0.05) | $G_{ENT, CA32DG}$ | 0.30 (0.19) |
| $G_{CA32DG, CA1}$ | 0.70 (0.02) | $G_{CA32DG, CA1}$ | 0.62 (0.03) |
| $G_{CA1, SUB}$ | 0.74 (0.03) | $G_{CA1, SUB}$ | 0.68 (0.07) |
| $G_{SUB, ENT}$ | 0.39 (0.04) | $G_{SUB, ENT}$ | 0.55 (0.07) |
| $G_{ENT, CA1}$ | 0.30 (0.08) | $G_{ENT, CA1}$ | 0.57 (0.08) |

*Table 3*. **Subgraphs Jaccard similarity index** for five pairs of regions of interest in the left medial temporal lobe, mean and standard deviation across 8 datasets of ten temporally ordered concatenated sessions. The Jaccard similarity index ranges from 0 to 1, where 1 indicates complete similarity of the subgraphs. Results for FASK and Two-Step.

*Connectivity subregions for pairs of regions: consistency across datasets*

From a subgraph $G_{P,Q}$ two subregions $S_P$ and $S_Q$ are obtained, one for each of the regions of interest, $P$ and $Q$. Two voxels $v_P \in V_P$ and $v_Q \in V_Q$ are members of subregion $S_P$ and $S_Q$, respectively, if there is an adjacency $v_P - v_Q$ in the subgraph $G_{P,Q}$. Formally, a subregion is defined as the set, $S_P = \{ v_P \in V_P \mid \text{the adjacency } v_P - v_Q \text{ exists in } A_{PQ} \}$, and conversely for $S_Q$. Put simply, a subregion $S_P$ for a pair of regions $P$ and $Q$, is a set containing voxels in $P$ that are connected to voxels in $Q$. Two subregions are computed for each of the five pairs of regions considered here. For example, from the subgraph $G_{ENT, CA32DG}$, the subregions $S_{ENT}$ and $S_{CA32DG}$ are obtained.

The Jaccard index can also be used to measure the similarity between subregions of communicating voxels across the eight datasets. For subregions the notation is adjusted as, $J(S_1, S_2) = |S_1 \cap S_2| / |S_1 \cup S_2|$, where $S_1$ and $S_2$ are subregions inferred from dataset $D_1$ and $D_2$ respectively. A value of 1 indicates that both subregions contain the exact same voxels. *Table 4* reports the mean Jaccard index for each subregion from each of the five pairs of regions considered, across the eight datasets. The results show that the two subregions for the pair (*ENT , CA32DG*) have the lowest mean Jaccard index, implying less consistency across datasets. The rest of the subregions show good consistency across datasets, indicating that FASK and Two-Step are recovering robust connectivity patterns that allow us to differentiate functional subregions driving the communication between larger regions of interest in the medial temporal lobe during rest.

| FASK | | | | Two-Step | | | |
|---|---|---|---|---|---|---|---|
| $S_{ENT}$ | 0.18 (0.10) | $S_{CA32DG}$ | 0.20 (0.12) | $S_{ENT}$ | 0.36 (0.20) | $S_{CA32DG}$ | 0.45 (0.16) |
| $S_{CA32DG}$ | 0.89 (0.02) | $S_{CA1}$ | 0.90 (0.02) | $S_{CA32DG}$ | 0.82 (0.03) | $S_{CA1}$ | 0.83 (0.03) |
| $S_{CA1}$ | 0.89 (0.03) | $S_{SUB}$ | 0.85 (0.03) | $S_{CA1}$ | 0.82 (0.05) | $S_{SUB}$ | 0.77 (0.06) |
| $S_{SUB}$ | 0.77 (0.04) | $S_{ENT}$ | 0.69 (0.04) | $S_{SUB}$ | 0.73 (0.07) | $S_{ENT}$ | 0.71 (0.07) |
| $S_{ENT}$ | 0.53 (0.08) | $S_{CA1}$ | 0.62 (0.06) | $S_{ENT}$ | 0.75 (0.09) | $S_{CA1}$ | 0.76 (0.13) |

*Table 4*. **Subregions Jaccard similarity index** for five pairs of regions of interest in the left medial temporal lobe, mean and standard deviation across 8 datasets of ten temporally ordered concatenated sessions. The Jaccard similarity index ranges from 0 to 1, where 1 indicates complete similarity of the communicating subregions. Results for FASK and Two-Step.

Subregions are defined only with adjacency information since they indicate voxels that are driving the connectivity, irrespectively of the direction, between two larger regions of interest. Nevertheless, given that FASK and Two-Step infer orientations of adjacencies, it is possible to estimate how many of the adjacencies in the subgraph $G_{P,Q}$ are in fact directed edges going from voxels in subregion $S_P$ to voxels in subregion $S_Q$ and how many go in the opposite direction. *Table 5* reports mean number of edges going in each of the directions for each of the subregions pairs across the eight datasets. These results show that on average, the medial temporal lobe subregions considered here have a similar number of edges going in both directions, suggesting strong reciprocal functional communication.

| | FASK | | | Two-Step | |
|---|---|---|---|---|---|
| | $\rightarrow$ | $\leftarrow$ | | $\rightarrow$ | $\leftarrow$ |
| $S_{ENT} - S_{CA32DG}$ | 5.75 (2.92) | 6.38 (2.26) | $S_{ENT} - S_{CA32DG}$ | 2.25 (1.39) | 4.50 (1.51) |
| $S_{CA32DG} - S_{CA1}$ | 101.38 (9.98) | 97.25 (6.98) | $S_{CA32DG} - S_{CA1}$ | 117.63 (6.28) | 121.00 (10.52) |
| $S_{CA1} - S_{SUB}$ | 24.00 (3.02) | 18.88 (2.85) | $S_{CA1} - S_{SUB}$ | 23.75 (3.85) | 25.25 (4.23) |
| $S_{SUB} - S_{ENT}$ | 29.00 (4.63) | 28.63 (4.75) | $S_{SUB} - S_{ENT}$ | 33.13 (0.83) | 32.25 (4.71) |
| $S_{ENT} - S_{CA1}$ | 7.38 (3.42) | 11.88 (4.36) | $S_{ENT} - S_{CA1}$ | 6.00 (1.85) | 6.63 (1.77) |

*Table 5.* **Number of directed edges** from one subregion to another for each of the five pairs of subregions from the left medial temporal lobe, mean and standard deviation across 8 datasets of ten temporally ordered concatenated sessions. Standard deviation in parentheses. For a row labeled as $S_P - S_Q$, the column labeled as $\rightarrow$ indicates mean number of directed edges from voxels in $S_P$ to voxels in $S_Q$, and conversely for the column $\leftarrow$ .

Finally, for illustration *Figures 2* and *Figure 3* show connectivity subregions in brain voxel space for each of the five pairs of subregions for FASK and Two-Step respectively, only for the first concatenated dataset. The figures for the rest of the seven datasets are similar (which can be inferred from the consistency results in *Table 4*). The regions are depicted in 3D voxels renderings and exploded views for better visualization. Voxels in the subregions are colored, and voxels not in the subregions are in gray. Since a region may receive signals from multiple regions and may also send signals to multiple regions, as the entorhinal cortex does, we might expect multiple subregions within such a region. Two (or more) subregions within a region of interest that communicate respectively with two (or more) subregions should have different, but possibly overlapping, sets of voxels. This can be seen clearly in *Figure 2* and *Figure 3*. These figures also show that the two algorithms, FASK and Two-Step, inferred similar subregions.

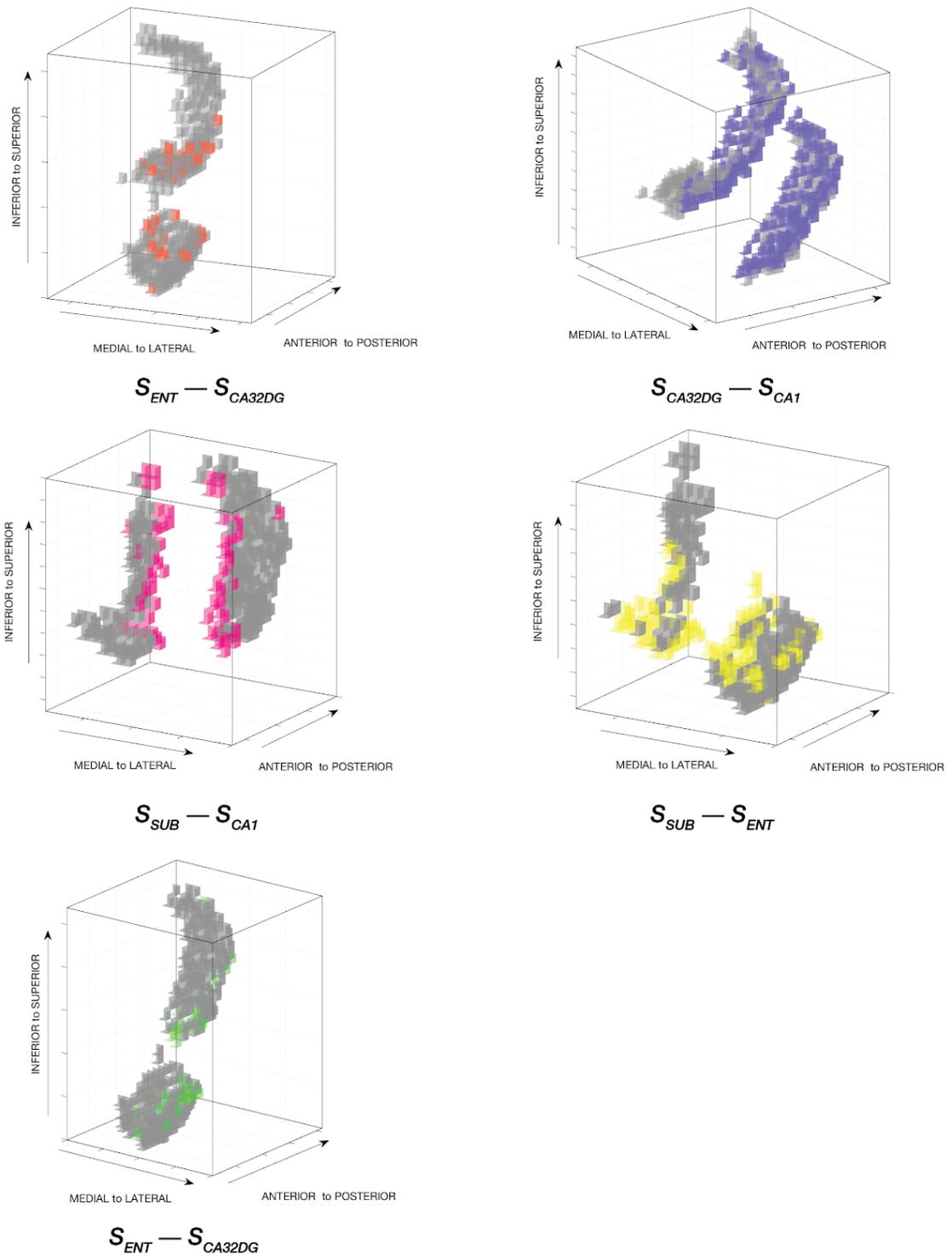

*Figure 2*. FASK, connectivity subregions for five pairs of regions of interest in the left hemisphere medial temporal lobe. Exploded maps of 3D renderings of regions of interest, indicating with colors the corresponding subregions. Orientation shown in each figure. Results for the first concatenated dataset.

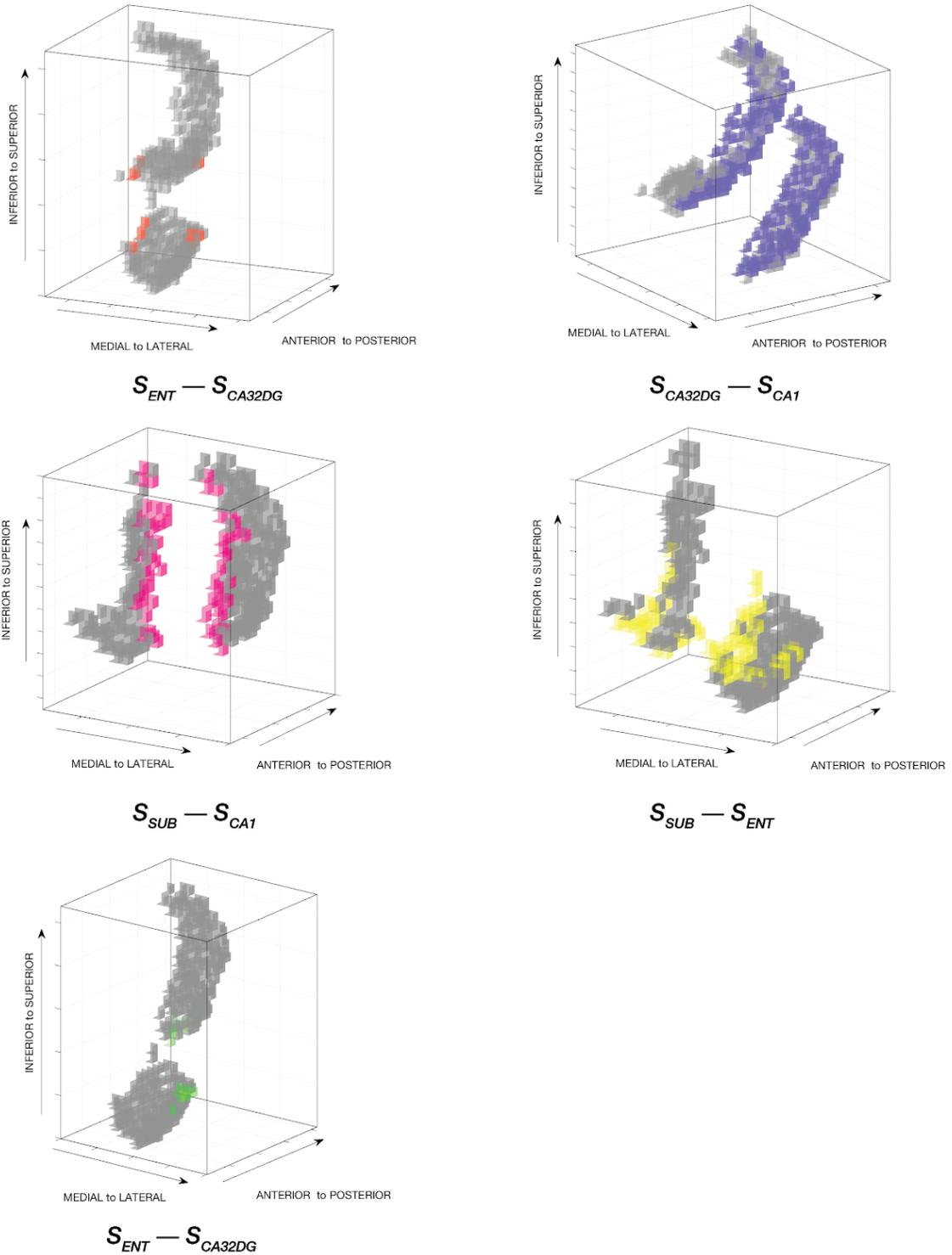

*Figure 3*. Two-Step, connectivity subregions for five pairs of regions of interest in the left hemisphere medial temporal lobe. Exploded maps of 3D renderings of regions of interest, indicating with colors the corresponding subregions. Orientation shown in each figure. Results for the first concatenated dataset.

**Discussion**

When voxel BOLD signals are clustered into regions of interest, it is common to seek connectivity relations between ROIs. The function of voxels within ROIs in that communication remains hidden. One might presume, for example, that voxels near the contact surfaces of anatomically distinct ROIs would be most involved in inter-ROI communication. Our experiments with the medial temporal lobe finds that such subregions are identifiable from signaling patterns between ROIs, without appeal to pre-specified locations within an ROI, and that they are overwhelming located near surfaces of regions in close contact. Our experiments also suggest that most inter-regional signaling in the medial temporal lobe is reciprocal.

While directions of influence are unstable between scans, adjacencies indicating an interaction among voxels in regions of the medial temporal lobe are quite stable.

Using simulated BOLD data for cyclic structures, Sanchez-Romero, et al. (2019) find that search results with FASK and Two-Step with resampling from a single scan, with smaller sample sizes, are comparable to those with concatenated scans such as those we have used here. This suggests that functional separations of subregions might be done with a single scan, but investigation of stability over multiple scans of the same subject seems advisable.

**Acknowledgments**

We thank M. R. K. Glymour, B. Huang, R. Poldrack and J.C. Liang. Research was supported by grants from the National Institutes of Health (1R01EB022858-01 FAIN – R01EB022858; 1R01LM012087 and 5U54HG008540-02  FAIN – U54HG008540). The content is solely the responsibility of the authors and does not necessarily represent the official views of the National Institutes of Health.

# Identification of Effective Connectivity Subregions

Ruben Sanchez-Romero, Joseph D. Ramsey, Kun Zhang, Clark Glymour
Department of Philosophy, Carnegie Mellon University, Pittsburgh, Pa., USA

**Supplementary Material**

Results for the right hemisphere medial temporal lobe data. Consistency across datasets *Tables S1 - S4*; directed edges across subregions *Table S5*; and 3D voxel space mapping of subregions for each pair of regions analyzed, *Figure S1* and *Figure S2*.

*Voxel level graph results: consistency across datasets*

Undirected Graphs

| Datasets | 1: (01, 10) | 2: (11, 20) | 3: (21, 30) | 4: (31, 40) | 5: (41, 50) | 6: (51, 60) | 7: (61, 70) | 8: (71, 80) |
|---|---|---|---|---|---|---|---|---|
| 1: (01, 10) | 1 | 0.811 | 0.784 | 0.798 | 0.784 | 0.803 | 0.797 | 0.814 |
| 2: (11, 20) |   | 1 | 0.787 | 0.803 | 0.802 | 0.801 | 0.811 | 0.805 |
| 3: (21, 30) |   |   | 1 | 0.79 | 0.783 | 0.785 | 0.778 | 0.79 |
| 4: (31, 40) |   |   |   | 1 | 0.802 | 0.812 | 0.808 | 0.802 |
| 5: (41, 50) |   |   |   |   | 1 | 0.789 | 0.792 | 0.787 |
| 6: (51, 60) |   |   |   |   |   | 1 | 0.809 | 0.792 |
| 7: (61, 70) |   |   |   |   |   |   | 1 | 0.793 |
| 8: (71, 80) |   |   |   |   |   |   |   | 1 |

Directed Graphs

| Datasets | 1: (01, 10) | 2: (11, 20) | 3: (21, 30) | 4: (31, 40) | 5: (41, 50) | 6: (51, 60) | 7: (61, 70) | 8: (71, 80) |
|---|---|---|---|---|---|---|---|---|
| 1: (01, 10) | 1 | 0.345 | 0.335 | 0.354 | 0.351 | 0.348 | 0.349 | 0.31 |
| 2: (11, 20) |   | 1 | 0.372 | 0.352 | 0.362 | 0.341 | 0.353 | 0.324 |
| 3: (21, 30) |   |   | 1 | 0.365 | 0.373 | 0.358 | 0.344 | 0.334 |
| 4: (31, 40) |   |   |   | 1 | 0.367 | 0.354 | 0.355 | 0.323 |
| 5: (41, 50) |   |   |   |   | 1 | 0.364 | 0.356 | 0.324 |
| 6: (51, 60) |   |   |   |   |   | 1 | 0.351 | 0.334 |
| 7: (61, 70) |   |   |   |   |   |   | 1 | 0.324 |
| 8: (71, 80) |   |   |   |   |   |   |   | 1 |

*Table S1*. **Jaccard similarity index for FASK** voxel level graphs from right medial temporal lobe data, one individual sampled across 80 sessions. Results indicate the Jaccard index for each pair of datasets, separately for undirected graphs and directed graphs. The Jaccard similarity index goes from 0 to 1, where 1 indicates complete similarity. Each dataset consists of 530 voxels corresponding to four regions of interest in the right hemisphere medial temporal lobe and ten temporally ordered sessions concatenated, resulting in 4610 datapoints. Each row/column indicates in parenthesis the range of the sessions that were concatenated for that particular dataset.

Undirected Graphs

| Datasets | 1: (01, 10) | 2: (11, 20) | 3: (21, 30) | 4: (31, 40) | 5: (41, 50) | 6: (51, 60) | 7: (61, 70) | 8: (71, 80) |
|---|---|---|---|---|---|---|---|---|
| 1: (01, 10) | 1 | 0.724 | 0.717 | 0.711 | 0.702 | 0.751 | 0.682 | 0.724 |
| 2: (11, 20) |   | 1 | 0.677 | 0.722 | 0.722 | 0.724 | 0.717 | 0.738 |
| 3: (21, 30) |   |   | 1 | 0.694 | 0.677 | 0.696 | 0.652 | 0.671 |
| 4: (31, 40) |   |   |   | 1 | 0.705 | 0.728 | 0.701 | 0.73 |
| 5: (41, 50) |   |   |   |   | 1 | 0.722 | 0.714 | 0.7 |
| 6: (51, 60) |   |   |   |   |   | 1 | 0.67 | 0.728 |
| 7: (61, 70) |   |   |   |   |   |   | 1 | 0.699 |
| 8: (71, 80) |   |   |   |   |   |   |   | 1 |

Directed Graphs

| Datasets | 1: (01, 10) | 2: (11, 20) | 3: (21, 30) | 4: (31, 40) | 5: (41, 50) | 6: (51, 60) | 7: (61, 70) | 8: (71, 80) |
|---|---|---|---|---|---|---|---|---|
| 1: (01, 10) | 1 | 0.394 | 0.39 | 0.374 | 0.367 | 0.425 | 0.355 | 0.394 |
| 2: (11, 20) |   | 1 | 0.374 | 0.397 | 0.392 | 0.407 | 0.409 | 0.407 |
| 3: (21, 30) |   |   | 1 | 0.382 | 0.359 | 0.372 | 0.335 | 0.352 |
| 4: (31, 40) |   |   |   | 1 | 0.378 | 0.433 | 0.405 | 0.395 |
| 5: (41, 50) |   |   |   |   | 1 | 0.411 | 0.392 | 0.367 |
| 6: (51, 60) |   |   |   |   |   | 1 | 0.351 | 0.4 |
| 7: (61, 70) |   |   |   |   |   |   | 1 | 0.381 |
| 8: (71, 80) |   |   |   |   |   |   |   | 1 |

*Table S2*. **Jaccard similarity index for Two-Step** voxel level graphs from right medial temporal lobe data, one individual sampled across 80 sessions. Results indicate the Jaccard index for each pair of datasets, separately for undirected graphs and directed graphs. The Jaccard similarity index goes from 0 to 1, where 1 indicates complete similarity. Each dataset consists of 530 voxels corresponding to four regions of interest in the right hemisphere medial temporal lobe and ten temporally ordered sessions concatenated, resulting in 4610 datapoints. Each row/column indicates in parenthesis the range of the sessions that were concatenated for that particular dataset.

*Subgraphs for pairs of regions: consistency across datasets*

| FASK | | Two-Step | |
|---|---|---|---|
| $G_{ENT, CA32DG}$ | 0.10 (0.10) | $G_{ENT, CA32DG}$ | 0.41 (0.17) |
| $G_{CA32DG, CA1}$ | 0.78 (0.03) | $G_{CA32DG, CA1}$ | 0.63 (0.03) |
| $G_{CA1, SUB}$ | 0.84 (0.04) | $G_{CA1, SUB}$ | 0.60 (0.07) |
| $G_{SUB, ENT}$ | 0.53 (0.07) | $G_{SUB, ENT}$ | 0.66 (0.06) |
| $G_{ENT, CA1}$ | 0.14 (0.07) | $G_{ENT, CA1}$ | 0.74 (0.30) |

*Table S3*. **Subgraphs Jaccard similarity index** for five pairs of regions of interest in the right medial temporal lobe, mean and standard deviation across 8 datasets of ten temporally ordered concatenated sessions. The Jaccard similarity index ranges from 0 to 1, where 1 indicates complete similarity of the subgraphs. Results for FASK and Two-Step.

*Connectivity subregions for pairs of regions: consistency across datasets*

| FASK | | | | Two-Step | | | |
|---|---|---|---|---|---|---|---|
| $S_{ENT}$ | 0.21 (0.17) | $S_{CA32DG}$ | 0.23 (0.16) | $S_{ENT}$ | 0.58 (0.20) | $S_{CA32DG}$ | 0.58 (0.20) |
| $S_{CA32DG}$ | 0.92 (0.02) | $S_{CA1}$ | 0.92 (0.01) | $S_{CA32DG}$ | 0.91 (0.02) | $S_{CA1}$ | 0.90 (0.02) |
| $S_{CA1}$ | 0.94 (0.02) | $S_{SUB}$ | 0.90 (0.02) | $S_{CA1}$ | 0.84 (0.04) | $S_{SUB}$ | 0.86 (0.04) |
| $S_{SUB}$ | 0.79 (0.05) | $S_{ENT}$ | 0.75 (0.06) | $S_{SUB}$ | 0.84 (0.06) | $S_{ENT}$ | 0.87 (0.05) |
| $S_{ENT}$ | 0.30 (0.13) | $S_{CA1}$ | 0.37 (0.09) | $S_{ENT}$ | 0.81 (0.21) | $S_{CA1}$ | 0.81 (0.21) |

*Table S4*. **Subregions Jaccard similarity index** for five pairs of regions of interest in the right medial temporal lobe, mean and standard deviation across 8 datasets of ten temporally ordered concatenated sessions. The Jaccard similarity index ranges from 0 to 1, where 1 indicates complete similarity of the communicating subregions. Results for FASK and Two-Step.

|  | FASK | | | Two-Step | |
| --- | --- | --- | --- | --- | --- |
|  | → | ← |  | → | ← |
| $S_{ENT}$ — $S_{CA32DG}$ | 2.63 (1.19) | 3.00 (1.41) | $S_{ENT}$ — $S_{CA32DG}$ | 1.13 (1.13) | 2.50 (1.77) |
| $S_{CA32DG}$ — $S_{CA1}$ | 104.75 (8.07) | 112.13 (8.89) | $S_{CA32DG}$ — $S_{CA1}$ | 118.88 (19.96) | 129.75 (12.08) |
| $S_{CA1}$ — $S_{SUB}$ | 19.75 (3.49) | 22.25 (2.38) | $S_{CA1}$ — $S_{SUB}$ | 47.25 (4.46) | 53.13 (6.69) |
| $S_{SUB}$ — $S_{ENT}$ | 16.50 (3.42) | 16.75 (1.98) | $S_{SUB}$ — $S_{ENT}$ | 13.75 (2.49) | 18.50 (3.63) |
| $S_{ENT}$ — $S_{CA1}$ | 3.00 (2.33) | 4.63 (0.92) | $S_{ENT}$ — $S_{CA1}$ | 0.38 (0.52) | 1.00 (0.53) |

*Table S5*. **Number of directed edges** from one subregion to another for each of the five pairs of subregions from the right medial temporal lobe, mean and standard deviation across 8 datasets of ten temporally ordered concatenated sessions. For a row labeled as $S_P$ — $S_Q$, the column labeled as → indicates mean number of directed edges from voxels in $S_P$ to voxels in $S_Q$, and conversely for the column ← .

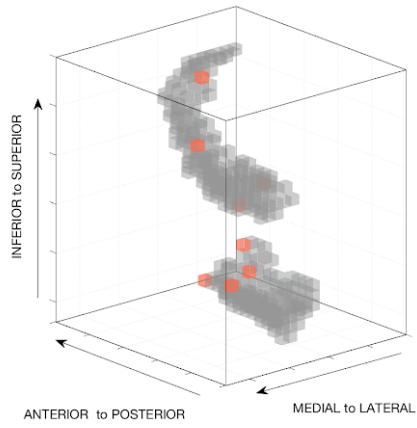

$S_{CA32DG} - S_{ENT}$

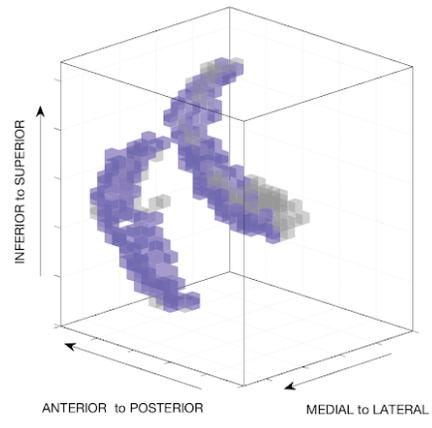

$S_{CA1} - S_{CA32DG}$

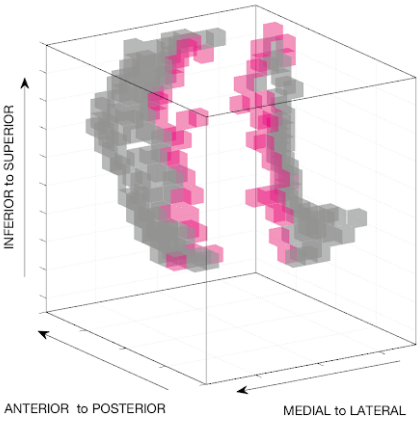

$S_{CA1} - S_{SUB}$

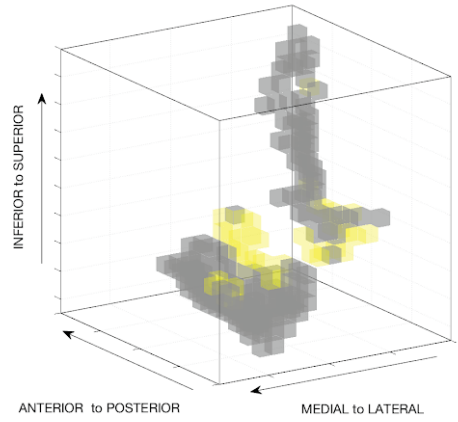

$S_{ENT} - S_{SUB}$

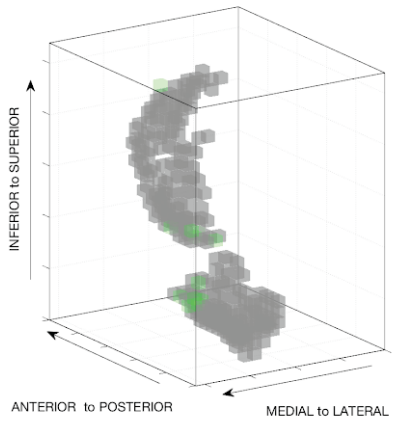

$S_{ENT} - S_{CA32DG}$

*Figure S1*. **FASK**, connectivity subregions for five pairs of regions of interest in the right hemisphere medial temporal lobe. Exploded maps of 3D renderings of regions of interest, indicating with colors the corresponding subregions. Orientation shown in each figure. Results for the first concatenated dataset.

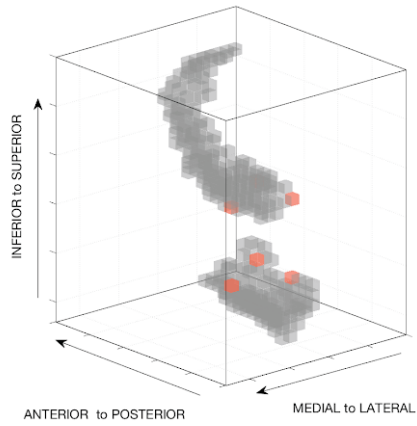

$S_{CA32DG} - S_{ENT}$

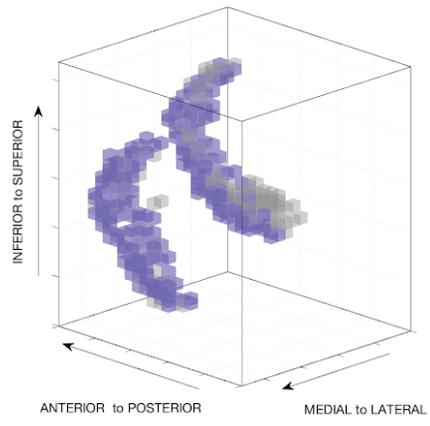

$S_{CA1} - S_{CA32DG}$

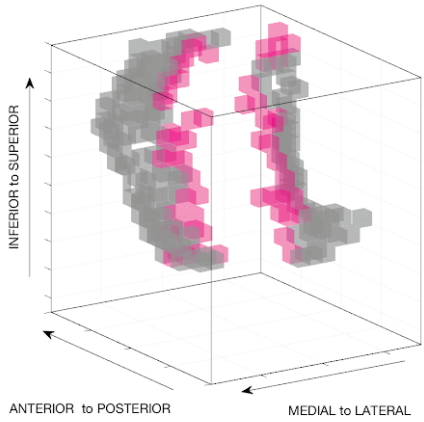

$S_{CA1} - S_{SUB}$

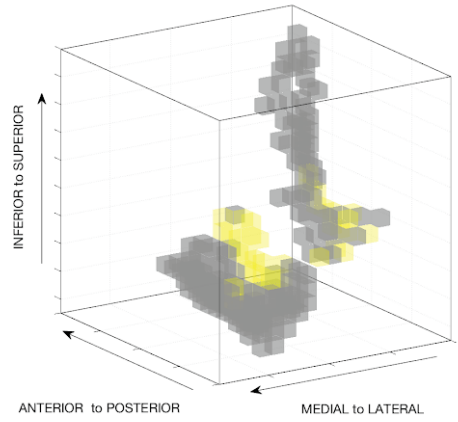

$S_{ENT} - S_{SUB}$

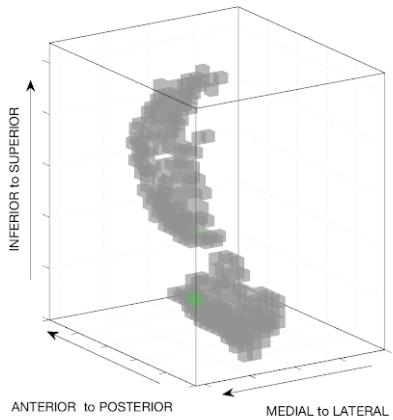

$S_{ENT} - S_{CA32DG}$

*Figure S2*. **Two-Step**, connectivity subregions for five pairs of regions of interest in the right hemisphere medial temporal lobe. Exploded maps of 3D renderings of regions of interest, indicating with colors the corresponding subregions. Orientation shown in each figure. Results for the first concatenated dataset.